*Review*

# Effect of Metals on Kinetic Pathways of Amyloid-β Aggregation

**Francis Hane [1] and Zoya Leonenko [1,2,\*]**

[1] Department of Biology, University of Waterloo, 200 University Ave, Waterloo, ON N2L 3G1, Canada; E-Mail: francishane@gmail.com (F.H.)

[2] Department of Physics and Astronomy, University of Waterloo, 200 University Ave, Waterloo, ON N2L 3G1, Canada

**\*** Author to whom correspondence should be addressed; E-Mail: zleonenk@uwaterloo.ca; Tel.: +1-519-888-4567 (ext. 38273); Fax: +1-519-746-8115.

**Abstract:** Metal ions, including copper and zinc, have been implicated in the pathogenesis of Alzheimer's disease through a variety of mechanisms including increased amyloid-β affinity and redox effects. Recent reports have demonstrated that the amyloid-β monomer does not necessarily travel through a definitive intermediary en-route to a stable amyloid fibril structure. Rather, amyloid-β misfolding may follow a variety of pathways resulting in a fibrillar end-product or a variety of oligomeric end-products with a diversity of structures and sizes. The presence of metal ions has been demonstrated to alter the kinetic pathway of the amyloid-β peptide which may lead to more toxic oligomeric end-products. In this work, we review the contemporary literature supporting the hypothesis that metal ions alter the reaction pathway of amyloid-β misfolding leading to more neurotoxic species.

**Keywords:** amyloid-metal effects; amyloid aggregation; multiple pathways kinetics; Alzheimer's disease

## 1. Introduction

Amyloid-β (Aβ), the protein implicated in Alzheimer's disease (AD) has been demonstrated to form a wide variety of structures including fibrils and a subset of smaller structures generically termed as amyloid oligomers [1]. Earlier research defined the term "oligomer" as a smaller pre-fibrillar structure lying on the kinetic pathway en-route to a mature fibril, however recent reports provide increasing evidence that these structures lie on independent and distinct kinetic pathways [2–4]. In addition, some species of oligomers, an end-product in and of themselves, have been demonstrated to convert to mature amyloid fibrils in the presence of small molecular catalysts [5].

An extensive amount of literature has been published on the effects of metal ions such as copper, zinc and iron, on Aβ aggregation and AD symptoms [6–12]. Metal ions have been demonstrated to alter the structure of amyloid aggregates and to inhibit fibril formation in some cases [13–17]. In addition, higher than physiological concentrations of metal ions have been found within the brain parenchyma and isolated from amyloid plaques [18].

An increasing amount of evidence indicates that the presence of metal ions alters the kinetic pathway of amyloid-β directing its aggregation away from a more stable fibrillar structure and towards a pathway resulting in more neurotoxic structures. This field of research is imperative to furthering pharmaceutical therapies which may, if not inhibit aggregation, at least direct aggregation along a less toxic pathway.

In this review, we cover the effect of metal ions on amyloid aggregate structure and the kinetics of aggregation and address how metal ions affect the reaction pathway of the amyloid aggregation process.

## 2. Fibrillar and Oligomeric Structures

### 2.1. The Amyloid Fibril

The amyloid fibril is intrinsic to the protein backbone accessible to all polypeptides: no primary sequence encodes for the fibril structure [1]. Despite early difficulties with solubility and crystallization, a number of groups have been able to determine the fibril structure using NMR and x-ray diffraction [19,20]. Mature fibrils of all proteins have been shown to be unbranched, approximately 10 nm in diameter composed of 2–4 nm wide protofilaments which wrap around one another to form the mature fibril [21]. Fibrils are composed of β-sheets which stack in a parallel or antiparallel form with the β-sheets perpendicular to the fibril axis [22] with hydrogen bonds parallel to the fibril axis holding the β-sheets together as shown in Figure 1 [21,23].

> **Figure 1.** Cartoon of β-sheet amyloid structure based on NMR spectra. Notice the two beta-sheets folding upon one another and stacked in a parallel manner. Figure reprinted with permission from Luhrs, *et al*. 2005). Copyright 2005, National Academy of Sciences, USA.

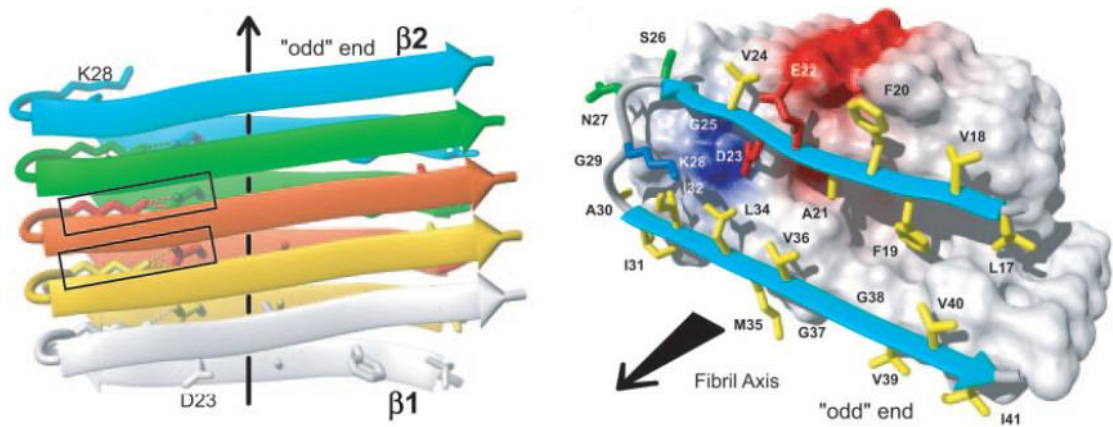

Nuclear magnetic resonance (NMR) experiments and molecular dynamics (MD) simulations have detected a salt bridge between Asp23 and Lys28 which is imperative for the self-dimerization of amyloid-β unto itself and an important element of fibril structure and further aggregation [24–27]. All fibrils have been shown to form a left-handed helix [28,29]. An increasing body of evidence has shown that the β-sheets within fibrils are arranged in a parallel conformation [30,31] although metastable anti-parallel stacking can occur as well [24]. These anti-parallel sheets may, however, eventually evolve towards a parallel conformation when in the presence of other parallel protofibrils [32].

*2.2. The Amyloid Oligomer*

Recent research has shown that oligomers are the neurotoxic species in a variety of neurodegenerative amyloid diseases [33,34]. The term "oligomer" is a broad term used to denote some form of non-fibrillar aggregates. The observed species identified as oligomers include disordered aggregates [33], micelles [35], protofibrils [36,37], prefibrillar aggregates [34], toxic amyloid-beta fibrillar oligomers (TABFO's) [38], amyloid diffusible ligands [39], prefibrillar oligomers (PFO's) [40], globulomers [41] and annular protofibrils (APF's) [42]. Perhaps the most interesting of these oligomeric structures is the membrane-associated ion channel referred to by Glabe as the APF which provides a direct biophysical mechanism explaining many AD symptoms. The APF has been demonstrated to act as a calcium ion channel affecting cellular homeostasis [43–46].

While this ever expanding list of amyloid species is fascinating, the discovery of new amyloid species has not been without criticism. Benilova *et al.* have argued that these species may in fact be "a way to explain inconsistencies in existing models without applying the scientific rigor needed to make real progress" [47]. In addition, it is very likely that many of these species, synthesized *in vitro* under very controlled conditions are not physiologically present and therefore hold little relevance for clinical Alzheimer's disease research.

Recently, Gu *et al.* were able to determine further details about the molecular structure of an amyloid-β oligomer using spin labeling [48]. Results from their work demonstrated that the amyloid hairpin can arrange itself in a variety of conformations (side-by-side and top-and-bottom) which may begin to explain the variety of oligomeric species that have been observed.

While an NMR or x-ray diffraction atomic level structure of amyloid-β has yet to be determined, Eisenberg and colleagues were recently able to obtain an x-ray diffraction structure of cylindrin which is similar to amyloid-β [49]. Their diffraction studies indicated that the cylindrin oligomer resembles a beta-barrel. While this work was conducted in solution, their observations tend to support the amyloid ion channels in lipid bilayers observed by Lal's group [44,46,50–52].

It has been also reported that the structure of oligomers and fibril formation itself is greatly affected by surfaces [53] as well as lipid membranes [54–56]. For example, protofibrils [53] (Figure 2), are unstable in solutions but can be stabilized by surfaces. Moores and others, [53,57,58] demonstrated that hydrophobic surface promotes the aggregation of amorphous structures, while charged surfaces promote the formation of protofibrils and fibrils. Hane and others [54,55,59,60] demonstrated that amyloid-β interactions with lipid membrane are defined by the lipid composition and charge and induce various defects in the membrane itself. Jang *et al.* demonstrated that multiple sub-units of amyloid-β fragments can form ion pores in the membrane which allow the passage of calcium ions leading to neuronal degeneration [61].

**Figure 2.** Diagram summarizing known amyloid-β aggregation pathways. The aggregation begins as an amyloid- β (Aβ) monomer which dimerizes eventually forming OC+ fibrillar oligomers (black pathway) [62]. The fibrillar oligomers polymerize to form mature amyloid fibrils [13]. Alternatively, the amyloid dimer can form A11+ prefibrillar oligomers (PFO) forming protofibrils (red pathway) [63,64]. These protofibrils may undergo an en-bloc conformational change to form amyloid fibrils [63]. The monomer may also travel along a pathway ending in amylospheriods (blue pathways) [4,65]. The pathways has a trimeric intermediate [66]. In the presence of copper (green pathway), amyloid dimerization is mediated by a copper ion forming small amyloid-copper oligomers [67] and eventually leading to larger amyloid-copper aggregates [13]. In the pathway mediated by lipid membranes (purple pathway), the amyloid dimer forms a hexameric ion pore [46] which may be identical to the annular protofibrils (APF) identified by the Glabe group [42] or the recent atomic structure of the amyloid oligomer [49]. These hexameric ion pores may stack to form deeper dodecameric structures [43]. This diagram was created using images from [4,13,40–43,49,62,63,66,67]. Images are reproduced with permission from the Nature Publishing Group, National Academy of Sciences, Public Library of Science, and the American Chemical Society.

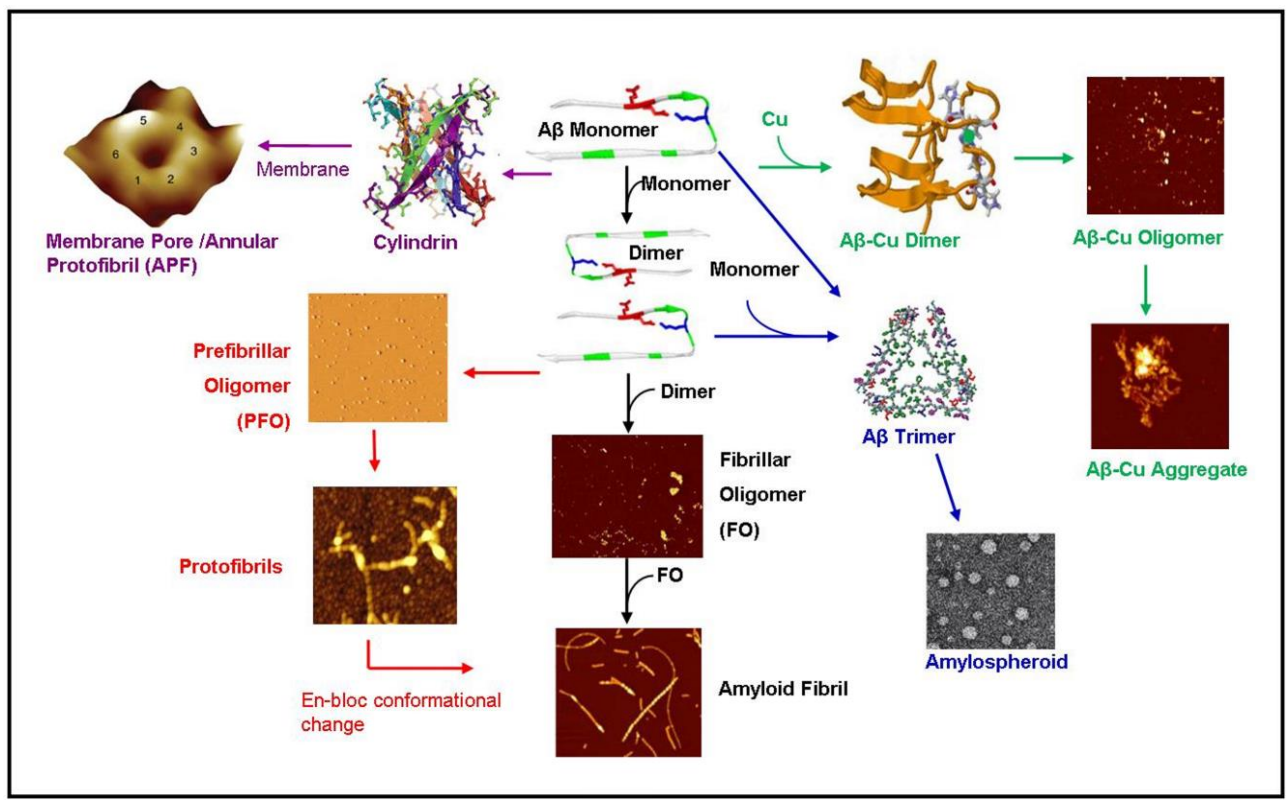

*2.3. Metal-Induced Amyloid Structures*

A considerable number of reports have provided empirical data showing aggregation structures of amyloid-β under the influence of metal ions, but a detailed NMR or x-ray diffraction atomic structure has yet to be reported [13,14]. However, Azimi *et al.* were able to use MD simulations to show that amyloid-β-copper coordinated structures can form in both an anti-parallel and parallel conformation. Copper ions have been shown to form an *intramolecular* complex while zinc ions tend to form *intermolecular* complexes cross linking multiple peptides [68–70]. The schematic in Figure 3 shows the details of amyloid-copper interactions [13]. Copper has been shown to coordinate with amyloid-β at the His13 and His14 site on one peptide with the His6 residue on the other peptide [11].

**Figure 3.** Cartoon of most probable amyloid-copper coordination. Notice the coordination sites at His6 from one peptide together with His13 and His14 from the second peptide. Figure reproduced with permission from [71]. Copyright American Chemical Society.

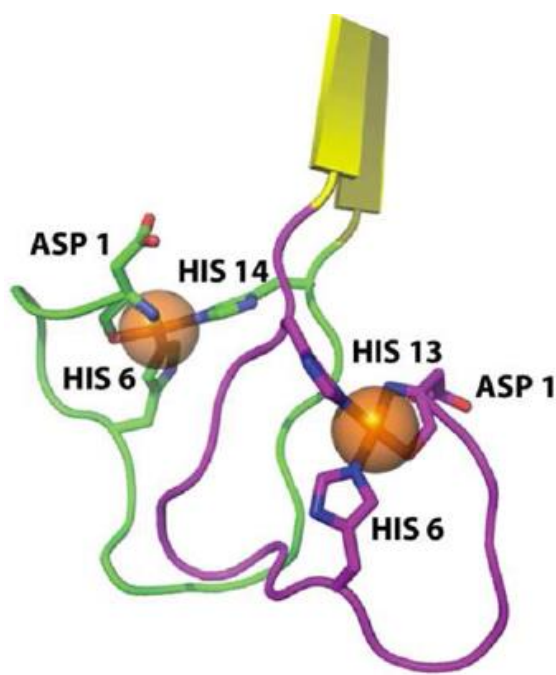

**3. Aggregation Kinetics, Thermodynamics and Aggregation Pathways**

Early amyloid research assumed that the kinetic pathway of the amyloid started as a monomer and progressed through the oligomer and protofibrillar structure en-route to a mature amyloid fibril. However, an increasing body of evidence suggests that these "intermediates" do not lie on the reaction pathway of the amyloid fibril but rather lie along a separate pathway. Demonstrating the complexity of amyloid kinetic pathways, we summarized the known pathways of amyloid-β as shown in Figure 2. The monomeric structure begins the aggregation pathway by dimerizing. The dimer can then follow a number of pathways. The dimer can follow a fibrillar oligomer pathway which is positive for the OC antibody leading to an amyloid fibril [62]. Alternatively, the dimer may be directed along the prefibrillar oligomer (PFO) pathway which is OC- but positive for the anti-oligomer antibody A11 [40] which leads to the production of protofibrils. The protofibrils may then undergo an "enbloc" conformational change to form amyloid fibrils [63]. In the presence of membrane lipids, the amyloid dimer may also form a hexameric ion pore which has been predicted and observed by Lal [46]. This structure may be identical to the hexameric structures observed by NMR by Laganowsky *et al*. [49] and observed by mass spectrometry by Bernstein *et al*. [43]. In the presence of copper, the amyloid-β dimerizes coordinated with copper ions and forms amyloid-copper aggregates [13,67]. Lastly, the amyloid monomer can form a trimer leading to amylospheroids [4,65].

Despite the apparent complexity in the aggregation pathways of amyloid-β, many of these oligomeric species were artificially induced *in vitro* under the influence of a variety of aggregation promoters or inhibitors. The precise aggregation pathway of amyloid-β in its physiological environment has yet to be elucidated.

*3.1.* In Vitro *Environment*

Work by the Glabe laboratory has used unique antibodies to study the divergence of these pathways. The antibodies OC and A11 have been used to study amyloid structures because they selectively bind different amyloid structures [72]. The fibrils have been shown to be OC+ and A11-, while the oligomers are OC- and A11+. These antibody studies have shown that prefibrillar oligomers (PFO) are A11+ while for fibrillar oligomer's (FO), the immediate precursors to mature fibrils are OC+ [40]. The difference in antibody binding is evidences of a structural difference between the two species. In an earlier study, Kayed *et al*. showed that some oligomers do not lie on the pathway to fibril formation [2,42]. Work by Hoshi and colleagues demonstrated that amylospheroids (ASPDs) lie off the pathway to fibrils and begin with trimerization as opposed to dimerization [4]. In addition, molecular dynamics simulations have revealed trimeric structures [11].

Molecular dynamics studies by Shea and colleagues revealed a "rich diversity" of aggregation pathways via a variety of mechanisms [73,74]. These mechanisms include the ordered assembly of oligomers into fibrils, the aggregation of non-fibrillar aggregates and the reorganization of amorphous aggregates into fibrils. Structures with the highest proportion of beta-sheet rich structures resulted in fibrils or barrel type structures, perhaps identical to those observed by Connoly *et al*. [46] and Lagonowky *et al*. [49]. Despite the higher beta-sheet content of barrel type structures, these never evolved into fibrils providing evidence that this structure lies on an alternate pathway to the amyloid fibril. In addition, an increase in the beta-sheet content of aggregates yielded fewer aggregation pathways indicating that beta-sheet structures stabilize the peptide and reduce the propensity for the peptide to fold into a disordered amorphous "glassy" state.

The pathway of amyloid-β aggregation appears to be concentration dependent displaying first order kinetics [75]. Lower concentrations tend to form PFO's while higher concentrations yield oligomers with lower membrane disruption and associated neurotoxicity, likely the result of increased hydrophobic segments [76].

*3.2. Copper Environment*

Metal ions, such as copper and zinc, have been demonstrated to modulate amyloid-β aggregation directing the aggregation pathways along different pathways [77,78]. Extensive work has been conducted on the effect of metals on the aggregation of amyloid-β all of which has shown that both Cu(II) and Zn(II) accelerate aggregation by shortening or eliminating the lag phase associated with the amyloid fibrillization process [8,69,79,80]. In the specific case of a copper rich environment, amyloid aggregation starts instantly [67]. Both copper and zinc ions have been shown to abolish the formation of fibrillar forms of amyloid-β in favour of amorphous precipitates especially at higher peptide concentrations [10,13,14] even though substoichiometric concentrations of amyloid-β have been shown to induce and accelerate the formation of amyloid fibrils [70]. However, other groups have reported an increase in the nucleation and elongation rates of fibril formation. These amyloid-copper fibril seeds have also been shown to initiate aggregation in non-copper solutions of amyloid-β [70]. It is unclear whether amorphous copper-amyloid-β fibril seeding complexes would still seed fibrils under previously studied experimental conditions or would seed amorphous aggregates. If the former is the case, that fibrils would be seeded by amorphous aggregates, it would likely provide evidence that an interconversion of the non-fibrillar to the fibrillar pathway can occur.

It is likely that the discrepancy between amorphous and fibrillar forms of the amyloid-complex can be attributed to the peptide and metal concentrations or preparation protocols. Work by Pedersen *et al*. showed that the Cu:Aβ ratio is a major determinant of the aggregation pathway [67]. Pedersen, together with other researchers identified three different kinetic pathways that amyloid-β may travel while under the influence of Cu(II) ions. The first pathway, where [Cu]<<[Aβ], occurs by the complex rapidly forming a critical nucleus with the slow elongation of the fibril as peptide-metal complexes are added to the nucleus [67,70,81]. At equimolar concentrations, a fast irreversible process dominates where peptide-copper oligomers slowly bind together to form amorphous aggregates and eventually spherical oligomers [67,80]. The third pathway where [Cu] > [Aβ] results in both fibrillar and oligomeric formation with higher copper concentrations resulting in higher proportions of oligomeric forms of amyloid-β likely the result of a destabilizing effect of copper on the amyloid-β structure [67]. It is believed that the aggregation of the copper-amyloid complex is rate limiting as opposed to the formation of the amyloid-copper nucleus which is the opposite of the nucleus formation rate limiting step in non-metallic amyloid-β aggregation [67].

*3.3. Zinc Environment*

Work by Bush *et al*. demonstrated that zinc destabilizes amyloid-β monomers and rapidly increased the rate of amorphous aggregates [81]. Further, Maggio and colleagues observed an approximate 40 fold increase in aggregation rate in zinc-amyloid-β solutions compared to solutions absent of zinc [82]. Similar to copper, zinc has been shown to coordinate with His13 and His14 which may drastically reduce the lag period allowing the aggregation process to essentially "bypass" the critical nucleation phase, which is often considered the rate limiting step in aggregation in non-metallic solutions [67]. Similar to copper, amyloid-β only forms amorphous aggregates in zinc environments [83]. The observation that these structures never convert to amyloid fibrils, even when providing conditions to overcome any possible kinetic barrier, provides evidence that amorphous aggregates lie on a separate kinetic pathway to amyloid fibrils.

Similar to copper and zinc, iron has also been shown to increase the aggregation rate of amyloid-β but not to the extent that zinc does [82].

*3.4. Physiological Environment*

Physiologically, the concentration of copper ions in the cerebrospinal fluid (CSF) is of a micromolar concentration which is far greater than the nanomolar CSF concentration of amyloid-β [84,85]. In extracellular fluid (ECF), amyloid-β concentrations decrease by an order of magnitude while copper concentrations are up to 100 times higher than in the CSF [86,87]. This difference in concentration results in fibrils tending to form in the ECF as opposed to the CSF [6]. These micromolar concentrations of copper are sufficient to induce amyloid-β aggregation providing an incentive for metal chelation therapies which attack the metallic mechanism for aggregation [10,80].

In contrast to copper, zinc ions are believed to promote aggregation, but the aggregates may not be neurotoxic leading some researchers to comment that some aggregation may in fact be a neuroprotective mechanism [88].

The aggregation process is driven by thermodynamics: once aggregation begins, there is no thermodynamic reason for the aggregation process to cease [89]. However, the aggregation of amyloid-β requires energy for the peptide to escape its energy well to add additional peptide to the fibril chain. As the peptide folds and aggregates, it travels along its free energy landscape along its kinetic pathway with oligomers or fibrils occupying local or global minima [90]. The dimerization of a monomer results in a significant reduction in the free energy of the system [91]. Parallel and antiparallel conformations result in different free energies [92]. The amyloid fibril has been shown to occupy the global free energy minima and is the most stable structure in most environments [89,93] with double- and triple-stranded structures being the most stable. Shea and colleagues demonstrated that barrel-like aggregates have similar potential energies as mature fibrils [74].

The equilibrium between oligomers and monomers has been shown to be a function of the peptide concentration to the power of the number of monomers per oligomer [94]. The addition of metals to amyloid solutions results in a higher proportion of antiparallel structure resulting in an approximately 25% decrease in potential energy compared to the parallel conformation [92]. The previously discussed change in aggregation behavior of amyloid-β in metal environments is likely caused by kinetic factors, not thermodynamic ones [95,96]. While a number of groups have reported copper-amyloid-β affinity and zinc-amyloid-β affinity [97–99], to our knowledge no group has yet reported amyloid-amyloid affinity in the presence of copper or zinc.

**4. Neurotoxicity of Metals**

In addition to their effect of increasing the aggregation rate of amyloid-β, metal ions have been universally acknowledged to contribute to oxidative stress and inflammation of the brain of Alzheimer's patients [6]. However oxidative stress also plays a role in normal aging [100,101]. Oxidative stress is one of the initial signs of AD [102] preceding the presence of inflammation and amyloid plaques. Oxidative stress is mediated by $H_2O_2$, which, *via* a Fenton reaction, produces the OH radical which is highly reactive and initiates a variety of reactions including post-translational protein modification, DNA damage and lipid peroxidation [18]. The brain does have natural defense mechanisms for neutralizing excess $H_2O_2$, but these become overwhelmed with the excessive amount of $H_2O_2$ and the highly reactive OH radical [103,104].

A number of reports have shown that both the lipid peroxidation and copper at the neuronal synapse promoting amyloid-β aggregation are a contributing factor of copper toxicity[78,105,106]. It would be expected that neurotoxicity would be a function of copper concentration. In fact, amyloid-β combined with nanomolar concentrations of copper is more toxic than amyloid-β with copper solutions which are 50 times more concentrated [70]. While Cu(II) ions have been shown to increase amyloid toxicity [107], there is some controversy in the literature as to whether Zn ions are neurotoxic or neuroprotective [83]. Upon review of the literature, it appears that at higher concentrations, Zn appears to have a neurotoxic effect whereas it has a neuroprotective effect at lower concentrations [83,107–109].

## 5. Conclusions

An overwhelming body of evidence demonstrates that the addition of trace metals considerably accelerates the kinetics of amyloid-β aggregation and may contribute to neurotoxicity as is the case of excess levels of copper ions. The diversity of amyloid stable structures which never evolve into amyloid fibrils, even when conditions to overcome a kinetic barrier is evidence that the amyloid aggregation process is not a single pathway with a variety of intermediaries, but rather many different pathways. Where these pathways diverge is still a matter of contemporary debate, but it likely occurs very early in the aggregation process, possibly even occurring at the initial dimerization of two monomers. Future research will need to focus on characterizing these oligomers and determining where these structures diverge.

## Acknowledgments

The authors acknowledge financial support from the Natural Sciences and Engineering Research Council of Canada (NSERC) and Melesa Hane for critical reading of the manuscript.

## Conflicts of Interest

The authors declare no conflict of interest.